\documentclass[11pt]{article}
\usepackage{amsmath,amsfonts,latexsym,graphicx,amssymb}
\date{}
\def\la{\langle\,}
\def\r{\,\rangle}
\def\<{\langle}
\def\>{\rangle}
\newcommand{\eeq}{\end{eqnarray}}
\newcommand{\beq}{\begin{eqnarray}}

\newcommand{\muu}{\mathfrak{u}}
\newcommand{\mf}{\mathfrak{f}}

\def\con{{}_{\_\rule{-1pt}{0pt}\_}
\rule{-2pt}{0pt}\raise1.5pt\hbox{$\mid$}\hspace{2pt}}

\newtheorem{theorem}{Theorem}
\newtheorem{proposition}{Proposition}

\title{\bf Quantum damped oscillator II:\\
Bateman's Hamiltonian vs. 2D
Parabolic Potential Barrier.}
\author{Dariusz Chru\'sci\'nski  \\
 Institute of Physics, Nicolaus Copernicus University \\
 ul. Grudzi\c{a}dzka 5/7, 87-100 Toru\'n, Poland}
\begin{document}

\maketitle

\begin{abstract}

We show that quantum Bateman's system which arises in the
quantization  of a damped harmonic oscillator is equivalent to a
quantum problem with 2D parabolic potential barrier known also as
2D inverted isotropic oscillator. It turns out that this system
displays the family of complex eigenvalues corresponding to the
poles of  analytical continuation of the resolvent operator to the
complex energy plane. It is shown that this representation is more
suitable than the hyperbolic one used recently by Blasone and
Jizba.

\end{abstract}

\vspace{.7cm}




\numberwithin{equation}{section}


\section{Introduction}
\setcounter{equation}{0}

In the previous paper \cite{I} we have investigated a quantization
of a 1D damped harmonic oscillator defined by the following
equation of motion
\begin{equation}\label{damped_osc}
 \ddot{x}+2\gamma \dot{x}+ \kappa x \;=\; 0\ ,
\end{equation}
where $\gamma>0$ denotes the damping constant. To quantize this
system we follow an old observation of Bateman \cite{Bat31} and
double the number of degrees of freedom, that is together with
(\ref{damped_osc}) we consider
\begin{equation}\label{damped_osc2}
 \ddot{y}-2\gamma \dot{y}+ \kappa y \;=\; 0\ ,
\end{equation}
i.e. an amplified oscillator. The detailed historical review of
the Bateman idea may be found in \cite{Dekker81}. For more recent
papers see e.g. \cite{Vit92} and \cite{Bla04}. The enlarged system
is a Hamiltonian one and it is governed by the following classical
Bateman Hamiltonian:
\begin{equation}\label{BHam}
    H(x,y,p_x,p_y)=
    p_x p_y -\gamma(xp_x - yp_y)+ \omega^2 xy\ ,
\end{equation}
where  $ \omega = \sqrt{\kappa - \gamma^2}\,
.$\footnote{Throughout the paper we shall consider the underdamped
case, i.e. $\kappa > \gamma^2$.} Now, performing a linear
canonical  transformation $(x,y,p_x,p_y) \longrightarrow
(x_1,x_2,p_1,p_2)$:
\begin{eqnarray}
x_1 & = &   \frac{p_y}{\sqrt{\omega}} \ , \hspace{1.5cm} p_1\, = \,-\sqrt{\omega}\, y \\
x_2 & = &  -\sqrt{\omega}\, x \ , \hspace{1cm} p_2 \,=\,
-\frac{p_x}{\sqrt{\omega}} \ ,
\end{eqnarray}
and applying a standard symmetric Weyl ordering one obtains the
following quantum Hamiltonian
\begin{equation}\label{H-old}
    \hat{H} = \omega\, \hat{\bf p} \wedge \hat{\bf x}  - \gamma \, \hat{\bf p} \odot\hat{\bf x}\  ,
\end{equation}
where $\hat{\bf x}=(\hat{x}_1,\hat{x}_2)$, $\hat{\bf
p}=(\hat{p}_1,\hat{p}_2)$ and we define two natural operations:
\[ \hat{\bf p} \wedge \hat{\bf x} = \hat{p}_1 \hat{x}_2 - \hat{p}_2 \hat{x}_1\ , \hspace{1cm}
\hat{\bf p} \odot \hat{\bf x} = \hat{\bf x} \odot \hat{\bf p} =
\frac 12\,\sum_{k=1}^2 ( \hat{x}_k \hat{p}_k + \hat{p}_k\hat{x}_k)
\ .
\]
Note, that $[\hat{\bf p} \wedge \hat{\bf x},\hat{\bf p}
\odot\hat{\bf x}] =0$.
 This operator was carefully analyzed in \cite{I}. In
particular it was shown that the family of complex eigenvalues
\begin{equation}\label{}
    \hat{H} |\mf^\pm_{nl} \> = E^\pm_{nl} |\mf^\pm_{nl}\>\ ,
\end{equation}
with
\begin{equation}\label{Enl}
    E^\pm_{nl} = \hbar\omega l \pm i \hbar \gamma (|l| + 2n +1)\ ,
\end{equation}
found already by Feshbach and Tikochinsky \cite{FT}, corresponds
to the poles of the resolvent operator $\hat{\rm R}(\hat{H},z) =
(\hat{H} - z)^{-1}$. Therefore, the corresponding generalized
eigenvectors $|\mf^\pm_{nl} \>$ may be interpreted as resonant
states of the Bateman system. It shows that dissipation of energy
is directly related to the presence of resonances.

In the present paper we continue to study this system but in a
different representation. Let us observe that  performing the
linear canonical transformation $({\bf x},{\bf p}) \longrightarrow
({\bf u},{\bf v})$:
\begin{equation}\label{CAN}
  {\bf x} = \frac{ \gamma {\bf u} - {\bf v}}{\sqrt{2\gamma}}\ , \ \ \ \ \
  {\bf p} = \frac{ \gamma {\bf u} + {\bf v}}{\sqrt{2\gamma}}\ ,
\end{equation}
one obtains for the Hamiltonian
\begin{equation}\label{H-new}
    \hat{H} = \omega\, \hat{\bf v}\wedge \hat{\bf u} + \hat{H}_{\rm iho}\ ,
\end{equation}
where
\begin{equation}\label{H-iho}
\hat{H}_{\rm iho} = \frac 12\,( \hat{\bf
    v}^2  -\gamma^2 \hat{\bf u}^2)\  ,
\end{equation}
represents a Hamiltonian of a 2D isotropic inverted harmonic
oscillator (iho) or, equivalently, a 2D potential barrier $-
\gamma^2 \hat{\bf u}^2$. Now, $\omega\, \hat{\bf v}\wedge \hat{\bf
u}$ generates an SO(2) rotation on $(u_1,u_2)$--plane. Therefore,
in the rotating frame the problem is described by the following
Schr\"odinger equation
\begin{equation}\label{}
    i\hbar\dot{\psi}_{\rm rf} = \hat{H}_{\rm iho}{\psi}_{\rm rf}\
    ,
\end{equation}
where the {\rm rotating frame} wave function ${\psi}_{\rm rf} =
\exp(i\omega\, \hat{\bf v}\wedge \hat{\bf u} t/\hbar)\, \psi$.

A 1D inverted (or reversed) oscillator was studied by several
authors in various contexts
\cite{Kemble,Wheeler,Friedman,Ann1,Ann2,Castagnino,Shimbori1}.
Recently, this system was studied in the context of dissipation in
quantum mechanics and a detailed analysis of its resonant states
was performed in \cite{damp2}. The present paper is mostly devoted
to analysis of a 2D iho.   We find its energy eigenvectors and
show that they are singular when one continues energy into complex
plane. The complex  poles correspond to resonant states of the 2D
potential barrier \cite{RES-1,RES-2}.

Finally, we analyze the Bateman system in the hyperbolic
representation used recently in \cite{Bla04} by Blasone and Jizba.
It turns out that this representation in not appropriate to
describe resonant states and hence the family of generalized
complex eigenvalues found in \cite{Bla04} is not directly related
to the spectral properties of the Bateman Hamiltonian. We stress
that it does not prove that these representation are physically
inequivalent. Clearly they are. Different representation lead to
different mathematical realization which is connected with
different functional spaces and different boundary conditions.
These may lead to different analytical properties and hence some
representation may display resonant states while others not.

From the mathematical point of view the natural language to
analyze the spectral properties of  Bateman's system is the so
called rigged Hilbert space approach to quantum mechanics
\cite{RHS1,RHS2,Bohm-Gadella,Bohm}.  We show (cf.
Section~\ref{ANALYTICITY}) that there are two dense subspaces
$\Phi_\pm \in L^2( \mathbb{R}^2_{\bf u})$ such that restriction of
the unitary group $\hat{U}(t) = e^{-i\hat{H}t/\hbar}$ to
$\Phi_\pm$ does no longer define a group but gives rise to two
semigroups: $\hat{U}_-(t)=\hat{U}(t)|_{\Phi_-}$ defined for $t\geq
0$ and $\hat{U}_+(t)=\hat{U}(t)|_{\Phi_+}$ defined for $t\leq 0$.
It means that the quantum  damped oscillator corresponds to the
following Gel'fand triplets:
\begin{equation}\label{}
  \Phi_\pm \, \subset \, L^2( \mathbb{R}^2_{\bf u}) \, \subset \, \Phi_\pm'\ ,
\end{equation}
and hence it serves as a simple example of Arno Bohm theory of
resonances \cite{Bohm}.

\section{2D inverted oscillator and complex eigenvalues}
\setcounter{equation}{0}

\subsection{2D harmonic oscillator}

Let us briefly recall the spectral properties of the 2D harmonic
oscillator (see e.g. \cite{Fluge,Kleinert}):
\begin{equation}\label{H-ho}
    \hat{H}_{\rm ho} = -\frac{\hbar^2}{2}\, \triangle_2 +
    \frac{\Omega^2}{2}\, \rho^2\ ,
\end{equation}
where the 2D Laplacian reads
\begin{equation}\label{2D-laplacian}
    \triangle_2 = \frac{\partial^2}{\partial\rho^2} + \frac{1}{\rho}
    \frac{\partial}{\partial\rho} + \frac{1}{\rho^2}
    \frac{\partial^2}{\partial\varphi^2}\ ,
\end{equation}
and $(\rho,\varphi)$ are standard polar coordinates on
$(u_1,u_2)$--plane. The corresponding eigenvalue problem
\begin{equation}\label{}
\hat{H}_{\rm ho} \psi_{nl}^{\rm ho} = \varepsilon_{nl}^{\rm
ho}\psi_{nl}^{\rm ho}\ ,
\end{equation}
  is solved by
\begin{equation}\label{}
\psi_{nl}^{\rm ho}(\rho,\varphi) = R_{nl}(\rho)\Phi_l(\varphi)\ ,
\end{equation}
where
\begin{equation}\label{}
    \Phi_l(\varphi) = \frac{e^{il\varphi}}{\sqrt{2\pi}}\ ,\ \ \ \ \ l=0,\pm1,\pm2,\ldots \ ,
\end{equation}
and the radial functions
\begin{equation}\label{}
R_{nl}(\rho) = C_{nl}\, (\sqrt{\Omega/\hbar}\, \rho)^{|l|}\,
\exp(-{\Omega}\rho^2/2\hbar)\,
{_1}F_1(-n,|l|+1,\Omega\,\rho^2/\hbar)\ ,
\end{equation}
where the normalization constant reads as follows
\begin{equation}\label{}
    C_{nl} = \frac{\sqrt{2\Omega/\hbar}}{|l|!}\, \sqrt{
    \frac{(n+|l|)!}{n!}}\ , \ \ \ \ \ \ n=0,1,2,\ldots\ .
\end{equation}
Finally, the corresponding eigenvalues $\varepsilon_{nl}^{\rm ho}$
are  given by the following formula
\begin{equation}\label{}
    \varepsilon_{nl}^{\rm ho} = \hbar \Omega ( |l| + 2n + 1)\ .
\end{equation}
Note, that using well known relation between confluent
hypergeometric function $_1F_1$ and generalized Laguerre
polynomials \cite{GR,AS}
\begin{equation}\label{L-1F1}
    L^\mu_n(z) = \frac{\Gamma(n+\mu +1)}{\Gamma(n+1)\Gamma(\mu
    +1)}\, _1F_1(-n,\mu+1,z)\ ,
\end{equation}
one may rewrite $R_{nl}$ alternatively as follows
\begin{equation}\label{}
R_{nl}(\rho) = \sqrt{2\Omega/\hbar}\,\sqrt{ \frac{n!}{(n+|l|)!}}
\, (\sqrt{\Omega/\hbar}\, \rho)^{|l|}\,
\exp(-{\Omega}\rho^2/2\hbar)\, L^{|l|}_n(\Omega\,\rho^2/\hbar)\ .
\end{equation}
It is evident that the family $\psi^{\rm ho}_{ln}$ is
orthonormal\begin{equation}\label{}
    \la \psi^{\rm ho}_{nl}| \psi^{\rm ho}_{n'l'}\r =
    \delta_{nn'}\delta_{ll'}\ ,
\end{equation}
and complete
\begin{equation}\label{}
  \sum_{n=0}^\infty \sum_{l=-\infty}^\infty\,
   \overline{\psi^{\rm ho}_{nl}(\rho,\varphi)}\, \psi^{\rm ho}_{nl}(\rho',\varphi') =
  \frac{1}{\rho}\,\delta(\rho-\rho')\delta(\varphi-\varphi')\ ,
\end{equation}
where $\la \ |\ \r$ denotes the standard scalar product in the
Hilbert space
\begin{equation}\label{}
    {\cal H} = L^2( \mathbb{R}_+,\rho d\rho) \otimes
    L^2([0,2\pi),d\varphi)\ .
\end{equation}

\subsection{Scaling and complex eigenvalues}

Let us note that  $\hat{H}_{\rm iho}$ defined in  (\ref{H-iho})
corresponds to the Hamiltonian of the harmonic oscillator with
purely imaginary frequency $\Omega = \pm i\gamma$. The connection
with a harmonic oscillator may be established by the following
scaling operator
\begin{equation}\label{V-lambda}
  \hat{V}_\lambda := \exp\left( \frac{\lambda}{\hbar}\
  \hat{\bf v} \odot \hat{\bf u}   \right)\ ,
 \end{equation}
with $\lambda \in \mathbb{R}$. Using commutation relation
$[\hat{u}_k,\hat{v}_l]=i\hbar\delta_{kl}$, this operator may be
rewritten as follows
\begin{equation}\label{}
 \hat{V}_\lambda =  e^{-i{\lambda}}\, \exp\left(-i\lambda\,
 \rho\frac{\partial}{\partial \rho}\right) \ ,
\end{equation}
and therefore it defines a complex dilation, i.e. the action of $
\hat{V}_\lambda$ on a function $\psi=\psi(\rho,\varphi)$ is given
by
\begin{equation}\label{}
\hat{V}_\lambda\, \psi(\rho,\varphi) = e^{-i{\lambda}}\, \psi(
e^{-i{\lambda}}\, \rho,\varphi)\ .
\end{equation}
In particular one easily finds:
\begin{equation}\label{}
  \hat{V}_\lambda\, \hat{H}_{\rm iho} \, \hat{V}_\lambda^{-1}
   = \frac 12\,e^{2i\lambda} \, \left( -{\hbar^2}\triangle_2 - e^{-4i\lambda}\gamma^2 \rho^2\right)\ .
\end{equation}
Therefore, for $e^{4i\lambda} = -1$, i.e. $\lambda = \pm \pi/4$,
one has
\begin{equation}\label{}
\hat{V}_{\pm \pi/4}\, \hat{H}_{\rm iho}  \, \hat{V}_{\pm
\pi/4}^{-1} = \pm i \,  \left( -\frac{\hbar^2}{2}\,\triangle_2 +
\frac{\gamma^2}{2}\, \rho^2\right)\ .
\end{equation}
Now, let us introduce
\begin{equation}\label{psi-psi-ho}
  \muu^{\pm}_{nl} = \hat{V}_{\mp \pi/4}\, \psi^{\rm ho}_{nl} \
  ,
\end{equation}
that is
\begin{equation}\label{}
\muu^{\pm}_{nl}(\rho,\varphi) = \sqrt{\pm i}\, \psi^{\rm
ho}_{nl}(\sqrt{\pm i}\rho,\varphi)\ .
\end{equation}
It is evident that
\begin{equation}\label{}
    \hat{H}_{\rm iho} \, \muu^{\pm}_{nl} = \varepsilon^\pm_{nl}\, \muu^{\pm}_{nl}\    ,
\end{equation}
where
\begin{equation}\label{}
\varepsilon^\pm_{nl} = \pm i\, \varepsilon^{\rm ho}_{nl} = \pm
i\hbar \gamma ( |l| + 2n + 1)\ .
\end{equation}
We stress that $\hat{V}_\lambda$ is not unitary (for $\lambda \in
\mathbb{R}$) and hence in general $\hat{V}_\lambda\psi$ does not
belong to $\cal H$ even for $\psi \in {\cal H}$. In particular the
generalized eigenvectors $\muu^\pm_{nl}$ do not belong to $\cal H$
(the radial part $R_{nl}(\sqrt{\pm i}\rho)$ is not an element from
$L^2( \mathbb{R}_+,\rho d\rho)$).

\begin{proposition} \label{PRO-psi}
Two families of generalized eigenvectors $\muu^\pm_{nl}$ satisfy
the following properties:
\begin{enumerate}


\item they are bi-orthonormal
\begin{equation}\label{}
    \int_0^{2\pi}\!\!\!\!\int_0^\infty\,
    \overline{\muu^\pm_{nl}(\rho,\varphi)}\,
     \muu^\mp_{n'l'}(\rho,\varphi)\,  \rho d\rho\,d\varphi = \delta_{nn'}\delta_{ll'}\ ,
\end{equation}

\item they are bi-complete
\begin{equation}\label{}
  \sum_{n=0}^\infty \sum_{l=-\infty}^\infty\,
   \overline{\muu^\pm_{nl}(\rho,\varphi)}\, \muu^\mp_{nl}(\rho',\varphi') =
  \frac{1}{\rho}\,\delta(\rho-\rho')\delta(\varphi-\varphi')\ .
\end{equation}
\end{enumerate}
\end{proposition}
The proof follows immediately from orthonormality and completness
of oscillator eigenfunctions $\psi^{\rm ho}_{nl}$.

\section{Spectral properties of the Bateman Hamiltonian}
\setcounter{equation}{0}

Now, we solve the corresponding spectral problem for the Bateman
Hamiltonian (\ref{H-new}). Note that $\hat{H}$ is bounded neither
from below nor from above and hence its spectrum $\sigma(\hat{H})
= (-\infty,\infty)$. The corresponding generalized eigenvectors
satisfy
\begin{equation}\label{}
    \hat{H} \psi_{\varepsilon,l} = E_{\varepsilon,l}
    \psi_{\varepsilon,l}\ ,
\end{equation}
where $l \in \mathbb{Z}$ and $\varepsilon \in \mathbb{R}$.
Assuming the following factorized form of $\psi_{\varepsilon,l}$
\begin{equation}\label{psi-el}
    \psi_{\varepsilon,l}(\rho,\varphi) = R_{\varepsilon,l}(\rho)
    \Phi_l(\varphi)\ ,
\end{equation}
one has
\begin{equation}\label{E-el}
    E_{\varepsilon,l} = \omega\hbar l + \varepsilon\ ,
\end{equation}
with
\begin{equation}\label{}
    \hat{H}_{\rm iho} R_{\varepsilon,l} = \varepsilon R_{\varepsilon,l}\ .
\end{equation}
The above equation rewritten in terms of
$(\rho,\varphi)$-variables takes the following form
\begin{equation}\label{}
\left( \frac{\partial^2}{\partial\rho^2} + \frac{1}{\rho}
    \frac{\partial}{\partial\rho} - \frac{|l|^2}{\rho^2}
    + \frac{\gamma^2}{\hbar^2}\, \rho^2 +
    \frac{2\varepsilon}{\hbar^2} \right) R_{\varepsilon,l} =0\ ,
\end{equation}
and its solution reads as follows
\begin{equation}\label{R-e}
    R_{\varepsilon,l}(\rho) = N_{\varepsilon,l}\,
    (\sqrt{i\gamma/\hbar}\,\rho)^{|l|}\, \exp(-i{\gamma}\rho^2/2\hbar)\,
{_1}\!F_1(a,|l|+1,i\gamma\rho^2/\hbar)\ ,
\end{equation}
with
\begin{equation}\label{a}
    a = \frac 12 \left( |l| + 1 - \frac{\varepsilon}{i\gamma\hbar}
    \right)\ .
\end{equation}
The normalization factor $N_{\varepsilon,l}$ is chosen such that
\begin{equation}\label{}
    \int_0^\infty \overline{R_{\varepsilon,l}(\rho)}\,
    R_{\varepsilon',l}(\rho)\, \rho d\rho =
    \delta(\varepsilon-\varepsilon')\ .
\end{equation}
It turns out (see Appendix A) that
\begin{equation}\label{N-el}
 N_{\varepsilon,l} = \sqrt{\frac{\gamma}{\pi |l|!}}\,
 (-i)^a\, \Gamma(a)\ ,
\end{equation}
with $a$ defined in (\ref{a}).

\begin{proposition} \label{psi-e-l}
The family of generalized eigenvectors $\psi_{\varepsilon,l}$
satisfy the following properties:
\begin{enumerate}
\item orthonormality
\begin{equation}\label{}
    \int_0^{2\pi}\!\!\!\!\int_0^\infty\,
    \overline{\psi_{\varepsilon,l}(\rho,\varphi)}\,
     \psi_{\varepsilon',l'}(\rho,\varphi)\,  \rho d\rho\,d\varphi = \delta(\varepsilon-\varepsilon')\delta_{ll'}\ ,
\end{equation}

\item completeness
\begin{equation}\label{}
 \sum_{l=-\infty}^\infty\,  \int_{-\infty}^\infty d\varepsilon\
   \overline{\psi_{\varepsilon,l}(\rho,\varphi)}\, \psi_{\varepsilon,l}(\rho',\varphi') =
  \frac{1}{\rho}\,\delta(\rho-\rho')\delta(\varphi-\varphi')\ .
\end{equation}
\end{enumerate}
\end{proposition}

Let us define another family of generalized energy eigenvectors
\begin{equation}\label{}
    \chi_{\varepsilon,l} = {\cal T}\psi_{\varepsilon,l}\ ,
\end{equation}
where the anti-unitary operator $\cal T$ is defined as follows
\begin{equation}\label{T-def}
{\cal T}\psi_{\varepsilon,l}(\rho,\varphi) =
\overline{R_{\varepsilon,l}(\rho)}\Phi_l(\varphi)\ .
\end{equation}
It easy to show that Bateman's Hamiltonian $\hat{H}$ is $\cal
T$--invariant
\begin{equation}\label{}
    {\cal T} \hat{H} {\cal T}^\dag = \hat{H}\ .
\end{equation}
Moreover, if $\psi(t) = \hat{U}(t)\psi_0$, then ${\cal T}\psi(t) =
\hat{U}(-t)({\cal T}\psi_0)$, which shows that $\cal T$ is a time
reversal operator. Finally, Proposition~\ref{psi-e-l} gives rise
to the following spectral representation of the Bateman
Hamiltonian
\begin{equation}\label{H-Psi}
    \hat{H} = \sum_{l=-\infty}^{+\infty} \int_{-\infty}^\infty\,
    d\varepsilon\, E_{\varepsilon,l} | \psi_{\varepsilon,l}\r \la
    \psi_{\varepsilon,l}|
= \sum_{l=-\infty}^{+\infty} \int_{-\infty}^\infty\,
    d\varepsilon\, E_{\varepsilon,l} | \chi_{\varepsilon,l}\r \la
    \chi_{\varepsilon,l}|
    \ ,
\end{equation}
with $E_{\varepsilon,l}$ defined in (\ref{E-el}).

\section{Analyticity, resolvent and resonances}
\setcounter{equation}{0} \label{ANALYTICITY}

Now, we continue energy eigenfunctions $\psi_{\varepsilon,l}$ and
$\chi_{\varepsilon,l}$ into complex $\varepsilon$-plane. Note,
that $\varepsilon$-dependence enters $R_{\varepsilon,l}$ via the
normalization factor $N_{\varepsilon,l}$ and the function
$_1F_1(a,|l|+1,i\gamma\rho^2/\hbar)$ ($a$ is
$\varepsilon$-dependent, see (\ref{a})).  It is well known (see
e.g. \cite{AS}) that confluent hypergeometric function
$_1\!F_1(a,b,z)$ defines a convergent series for all values of
complex parameters $a$, $b$ and $z$ provided $a\neq -n$ and $b\neq
-m$, with $m$ and $n$ positive integers. Moreover, if $a=-n$ and
$b\neq -m$, then $_1\!F_1(a,b,z)$ is a polynomial of degree $n$ in
$z$. In our case $b=|l|+1$ which is never negative and hence
$_1F_1(a,|l|+1,i\gamma\rho^2/\hbar)$ is analytic in $\varepsilon$.
However, it is no longer true for the normalization constant
$N_{\varepsilon,l}$ given by (\ref{N-el}). The $\Gamma$-function
has simple poles at $a=-n$, with $n=0,1,2,\ldots$, which
correspond to
\begin{equation}\label{e-n-l}
    \varepsilon = \varepsilon_{nl} = i\gamma \hbar(|l| + 2n +1)\ ,
\end{equation}
on the complex $\varepsilon$-plane. On the other hand the
time-reversed function $ \overline{R_{\varepsilon,l}}$ has simple
poles at $\varepsilon=\overline{\varepsilon_{nl}} =
-\varepsilon_{nl}$.

It is, therefore, natural to introduce two  classes of functions
that respect these analytical properties of $\psi_{\varepsilon,l}$
and $\chi_{\varepsilon,l}$. Recall \cite{Duren} that a smooth
function $f=f(\varepsilon)$ is in the Hardy class from above
${\cal H}^2_+$ (from below ${\cal H}^2_-$) if $f(\varepsilon)$ is
a boundary value of an analytic function in the upper, i.e.
$\mbox{Im}\, \varepsilon\geq 0$ (lower, i.e. $\mbox{Im}\,
\varepsilon\leq 0$) half complex $\varepsilon$-plane vanishing
faster than any power of $\varepsilon$ at the upper (lower)
semi-circle $|\varepsilon| \rightarrow \infty$. Define
\begin{equation}
\Phi_- := \Big\{ \phi \in {\cal S}( \mathbb{R}^2_{\bf u})\, \Big|
\, f(\varepsilon):= \la \chi_{\varepsilon,l} | \phi \r \in {\cal
H}^2_-\, \Big\} \ ,
\end{equation}
and
\begin{equation}
\Phi_+ := \Big\{ \phi \in {\cal S}( \mathbb{R}^2_{\bf u})\, \Big|
\, f(\varepsilon):=\la \psi_{\varepsilon,l} | \phi \r \in {\cal
H}^2_+\, \Big\} \ ,
\end{equation}
where ${\cal S}( \mathbb{R}^2_{\bf u})$ denotes the Schwartz space
\cite{Yosida}, i.e. the space of $C^\infty(\mathbb{R}^2_{\bf u})$
functions $f=f(u_1,u_2)$ vanishing at infinity ($|{\bf u}|
\longrightarrow \infty$) faster than any polynomial.

 It is evident from (\ref{T-def}) that
\begin{equation}\label{}
  \Phi_+ = {\cal T}({\Phi_-})\ .
\end{equation}
The main result of this section consists in the following
\begin{theorem}\label{MAIN} For any function $\phi^\pm \in \Phi_\pm$ one has
\beq   \label{phi+} \phi^+ =
\sum_{n=0}^\infty\sum_{l=-\infty}^\infty \muu^+_{nl} \la
\muu^-_{nl}|\phi^+\r \ , \eeq and \beq  \label{phi-} \phi^- =
\sum_{n=0}^\infty\sum_{l=-\infty}^\infty \muu^-_{nl} \la
\muu^+_{nl}|\phi^-\r \ . \eeq
\end{theorem}
For the proof see Appendix B.  The above theorem implies the
following spectral resolutions of the Hamiltonian:
\begin{equation} \label{GMH-1}
\hat{H}_-  \equiv \hat{H}\Big|_{\Phi_-} \, = \, \sum_{n=0}^\infty
\sum_{l=-\infty}^\infty\, {E^-_{nl}}\,
|\muu^-_{nl}\>\<\muu^+_{nl}|\ ,
\end{equation}
 and
\begin{equation} \label{GMH-2}
\hat{H}_+   \equiv \hat{H}\Big|_{\Phi_+} \, = \,
  \sum_{n=0}^\infty
\sum_{l=-\infty}^\infty\, {E^+_{nl}}\,
|\muu^+_{nl}\>\<\muu^-_{nl}|\ .
\end{equation}
 In the above formulae $E^\pm_{nl}$ is given by (\ref{Enl}).

 The same techniques may be applied for the resolvent
operator
\begin{equation}\label{}
  \hat{R}(z,\hat{H}) = \frac{1}{\hat{H}-z}\ .
\end{equation}
One obtains
\begin{eqnarray}\label{}
\hat{R}_+(z,\hat{H})  &=& \sum_{l=-\infty}^\infty
\int_{-\infty}^\infty\, \frac{d\varepsilon}{E_{\varepsilon,l}-z}\,
|\psi_{\varepsilon,l}\>  \< \psi_{\varepsilon,l}|\
\Big|_{\Phi_+}\nonumber \\ &=&
\sum_{n=0}^\infty\sum_{l=-\infty}^\infty \frac{1}{E^+_{n,l}-z}\,
|\muu^-_{nl}\> \la \muu^+_{nl}|\ ,
\end{eqnarray}
on $\Phi_+$, and
\begin{eqnarray}\label{}
\hat{R}_-(z,\hat{H})  &=& \sum_{l=-\infty}^\infty
\int_{-\infty}^\infty\, \frac{d\varepsilon}{E_{\varepsilon,l}-z}\,
|\chi_{\varepsilon,l}\>  \< \chi_{\varepsilon,l}|\
\Big|_{\Phi_-}\nonumber \\ &=&
\sum_{n=0}^\infty\sum_{l=-\infty}^\infty \frac{1}{{E^-_{n,l}}-z}\,
|\muu^+_{nl}\> \la \muu^-_{nl}|\ ,
\end{eqnarray}
 on $\Phi_-$. Hence,
$\hat{R}_+(z,\hat{H})$ has poles at $z=E^+_{nl}$, and
$\hat{R}_-(z,\hat{H})$ has poles at $z={E^-_{nl}}$. As usual
eigenvectors $\muu^+_{nl}$ and $\muu^-_{nl}$ corresponding to
poles of the resolvent are interpreted as resonant states. Note,
that the Cauchy integral formula implies
\begin{equation}\label{}
\hat{P}^+_{nl} \, :=\, |\muu^-_{nl}\r\la\muu^+_{nl}| \, =\,
\frac{1}{2\pi i} \oint_{\Gamma^+_{nl}} \hat{R}_+(z,\hat{H})dz
   \ ,
\end{equation}
where $\Gamma^+_{nl}$ is a clockwise closed curve that encircles
the singularity $z=E^+_{nl}$. Similarly,
\begin{equation}\label{}
\hat{P}^-_{nl} \, :=\, |\muu^+_{nl}\r\la\muu^-_{nl}| \, =\,
\frac{1}{2\pi i} \oint_{\Gamma^-_{nl}} \hat{R}_-(z,\hat{H})dz
   \ ,
\end{equation}
where $\Gamma^-_{nl}$ is an anti-clockwise closed curve that
encircles the singularity $z=E^-_{nl}$. One easily shows that
\begin{equation}\label{}
  \hat{P}^\pm_{nl} \cdot \hat{P}^\pm_{n'l'} = \delta_{nn'}\delta_{ll'}\,\hat{P}^\pm_{nl}\ ,
\end{equation}
and hence the spectral decompositions of (\ref{GMH-1}) and
(\ref{GMH-1}) may be written as follows:
\begin{equation}\label{H+-}
  \hat{H}_\pm =  \sum_{n=0}^\infty\sum_{l=-\infty}^\infty \, E^\pm_{nl}\, \hat{P}^\pm_{nl} \ .
\end{equation}
Finally, let us note, that restriction of the unitary group
$\hat{U}(t)=e^{-i\hat{H}t/\hbar}$  to $\Phi_\pm$ no longer defines
a group. It gives rise to two semigroups:
\begin{equation}
 \hat{U}_-(t)\, :=\, e^{-i\hat{H}_-t/\hbar} \ :\ \Phi_-
\ \longrightarrow\ \Phi_-\ ,\ \ \ \ \ \ {\rm for}\ \ \ t\geq 0\ ,
\end{equation}
and
\begin{equation}
  \hat{U}_+(t) \, :=\, e^{-i\hat{H}_+t/\hbar} \ :\ \Phi_+ \ \longrightarrow\ \Phi_+\ ,\ \ \ \ \ \
{\rm for}\ \ \ t\leq 0\ .
\end{equation}
Using (\ref{H+-}) and the formula for $E_{nl}$ one finds:
\begin{eqnarray} \label{phi:-}
\phi^-(t) = \hat{U}_-(t)\phi^- = \sum_{l=-\infty}^\infty
e^{-i\omega lt}\, \sum_{n=0}^\infty e^{-\gamma(2n + |l| +1)t} \,
\hat{P}^-_{nl} \ ,
\end{eqnarray}
for $t\geq 0$, and
\begin{eqnarray} \label{phi:+}
\phi^+(t) = \hat{U}_+(t)\phi^+ = \sum_{l=-\infty}^\infty
e^{-i\omega lt}\, \sum_{n=0}^\infty e^{\gamma(2n + |l| +1)t} \,
\hat{P}^+_{nl}\ ,
\end{eqnarray}
for $t\leq 0$.  We stress that $\phi^-_t$ ($\phi^+_t$) does belong
to $L^2(\mathbb{R}^2_{\bf u} )$ also for $t<0$ ($t>0$). However,
$\phi^-_t \in \Phi_-$ ($\phi^+_t \in \Phi_+$) only for $t\geq 0$
($t\leq 0$). This way the irreversibility enters the dynamics of
the reversed oscillator by restricting it to the dense subspace
$\Phi_\pm$ of $L^2(\mathbb{R}^2_{\bf u})$.

From the mathematical point of view the above construction gives
rise to so called rigged Hilbert spaces (or Gel'fand triplets)
\cite{RHS1,RHS2,Bohm-Gadella,Bohm}:
\begin{equation}\label{}
    \Phi_- \subset {\cal H} \subset \Phi_-'\ ,
\end{equation}
and
\begin{equation}\label{}
    \Phi_+ \subset {\cal H} \subset \Phi_+'\ ,
\end{equation}
where $\Phi_\pm'$ denote  dual spaces, i.e. linear functionals on
$\Phi_\pm$. Note, that generalized eigenvectors $\muu^\pm_{nl}$
are not elements from $\cal H$. However, they do belong to
$\Phi_\pm'$. The first triplet  $(\Phi_-,{\cal H},\Phi_-')$ is
corresponds to the evolution for $t\geq 0$, whereas the second one
$(\Phi_+,{\cal H},\Phi_+')$ corresponds to the evolution for
$t\leq 0$.

\section{Bateman's system in hyperbolic representation}
\label{Hyperbolic} \setcounter{equation}{0}

In a recent paper \cite{Bla04}  Blasone and  Jizba used another
representation. They transform Bateman's Hamiltonian (\ref{BHam})
into the following form
\begin{equation}\label{}
    H(y_1,y_2,w_1,w_2) = \frac 12 (w_1^2 - w_2^2) - \gamma( y_1w_2
    + y_2w_1) + \frac 12 \omega^2 (y_1^2 - y_2^2) \ ,
\end{equation}
with the new positions
\begin{equation}\label{}
    y_1 = \frac{x+y}{\sqrt{2}}\ , \hspace{1cm}  y_2 = \frac{x-y}{\sqrt{2}}\ ,
\end{equation}
and new canonical momenta
\begin{equation}\label{}
    w_1 = \frac{p_x+p_y}{\sqrt{2}}\ , \hspace{1cm}  w_2 = \frac{p_x-p_y}{\sqrt{2}}\
    .
\end{equation}
Now, introducing hyperbolic coordinates $(\varrho,u)$:
\begin{equation}\label{}
    y_1 = \varrho \, \cosh u \ , \hspace{1cm} y_2 = \varrho \, \sinh u \ ,
\end{equation}
the canonical quantization leads to the following Hamiltonian
defined on the Hilbert space ${\cal H} = L^2( \mathbb{R}_+,\varrho
d\varrho) \otimes L^2( \mathbb{R},du)$:
\begin{equation}\label{HB-hyper}
    \hat{H} = \hat{H}_0 + \hat{H}_{\rm iho}\ ,
\end{equation}
with
\begin{equation}\label{H0}
    \hat{H}_0 = - \frac{\hbar^2}{2}\, \square_{\,2} +
    \frac{\omega^2}{2}\, \varrho^2\ ,
\end{equation}
and the iho part $\hat{H}_{\rm iho}$
\begin{equation}\label{}
    \hat{H}_{\rm iho}= i\gamma\hbar\, \frac{\partial}{\partial u}\ .
\end{equation}
In the above formulae $\square_{\,2}$ denotes the 2D wave
operator, that is
\begin{equation}\label{}
    \square_{\,2} = \frac{\partial^2}{\partial y_1^2} - \frac{\partial^2}{\partial
    y_2^2}\, =\, \frac{\partial^2}{\partial\varrho^2} + \frac{1}{\varrho}
    \frac{\partial}{\partial\varrho} - \frac{1}{\varrho^2}
    \frac{\partial^2}{\partial  u^2}\ .
 \end{equation}
Clearly, in the $(\varrho,u)$ variables the formula for
$\hat{H}_{\rm iho}$ considerably simplifies and $\hat{H}_{\rm
iho}$ represents the generator of SO(1,1) hyperbolic rotation on
the $(y_1,y_2)$--plane. In this particular representation
$\hat{H}_{\rm iho}$ defines a self-adjoint operator on $L^2(
\mathbb{R},du)$. The corresponding eigen-problem is immediately
solved
\begin{equation}\label{}
    \hat{H}_{\rm iho} \Phi_\nu = \gamma \hbar\nu \, \Phi_\nu\ ,
\end{equation}
with $\Phi_\nu(u) = {e^{-i\nu u}}/{\sqrt{2\pi}}\ $, and hence it
reproduces the continuous spectrum of 2D iho $ \sigma(
\hat{H}_{\rm iho}) = (-\infty,\infty)$. However, there is a
crucial difference between elliptic $(\rho,\varphi)$ and
$(\varrho,u)$ representations. The generalized eigenvectors
$\Phi_\nu$ may be analytically continued on the entire complex
$\nu$--plane. Therefore, the hyperbolic representation does not
display the family of resonances corresponding to complex
eigenvalues $\varepsilon_{nl}$ defined in (\ref{e-n-l}). Of course
one may by hand fix the values of $\nu$ to $\nu = i(2n + |l| + 1)$
but then the corresponding discrete $\Phi_{nl}$ family is neither
bi-orthogonal nor bi-complete (cf. Proposition~\ref{psi-e-l}).

To show how the complex eigenvalues of   Blasone and Jizba
\cite{Bla04} appear let us consider $\hat{H}_0$ defined in
(\ref{H0}). Note, that $\hat{H}_0$  resembles 2D harmonic
oscillator given by (\ref{H-ho}). There is, however, crucial
difference between $\hat{H}_0$ and $\hat{H}_{\rm ho}$. The
hyperbolic operator `$-\square_{\,2}$', contrary to the elliptic
one `$-\triangle_2$', is not positively defined and hence it
allows for negative eigenvalues. It is clear, since in the
elliptic $(\rho,\varphi)$--representation $\hat{H}_0 =
i\omega\hbar\,
\partial_\varphi$ defines a self-adjoint operator on
$L^2([0,2\pi),d\varphi)$ with purely discrete spectrum
$\omega\hbar l$  ($l \in \mathbb{Z}$). Now,  the spectral analysis
of the Bateman Hamiltonian represented by (\ref{HB-hyper}) is
straightforward:
\begin{equation}\label{}
    \hat{H}\psi_{\epsilon\nu} = {\cal E}_{\epsilon\nu}\psi_{\epsilon\nu}
    \ ,
\end{equation}
with
\begin{equation}\label{}
    {\cal E}_{\epsilon\nu} = \epsilon + \gamma\hbar\nu\ ,
\end{equation}
and the following factorized form of $\psi_{\epsilon\nu}$:
\begin{equation}\label{psi-e-nu}
\psi_{\epsilon\nu}(\varrho,u) = {\cal R}_{\epsilon\nu}(\varrho)\,
\Phi_\nu(u)\ .
\end{equation}
 The radial function ${\cal R}_{\epsilon\nu}$ solves
\begin{equation}\label{}
    \hat{H}_0 {\cal R}_{\epsilon\nu} = \epsilon\, {\cal
    R}_{\epsilon\nu}\ ,
\end{equation}
and in analogy to (\ref{R-e}) it is given by
\begin{equation}\label{R-U}
    {\cal R}_{\epsilon\nu}(\varrho) = N_{\epsilon\nu} \,
    (\sqrt{\omega/\hbar}\varrho)^{i\nu} \exp(
    -\omega\varrho^2/2\hbar)\, U(b,i\nu + 1,\omega\varrho^2/\hbar)\ ,
\end{equation}
with
\begin{equation}\label{}
    b = \frac 12 \left( i\nu + 1 - \frac{\epsilon}{\hbar\omega}
    \right)\ .
\end{equation}
In (\ref{R-U}) we have used instead of the standard confluent
hypergeometric function $_1F_1$ so called Tricomi function $U$
(see e.g. \cite{BE}).\footnote{Actually, in  \cite{BE} (and also
in \cite{LL}) this function is denoted by $G$. We follow the
notation of Abramowitz and Stegun \cite{AS}.} It is defined by
\begin{equation}\label{}
    U(a,c,z) = \frac{\Gamma(1-c)}{\Gamma(a-c+1)}\, _1F_1(a,c,z) +
    \frac{\Gamma(c-1)}{\Gamma(a)}\, z^{1-c}\, _1F_1(a-c+1,2-c,z) \
    .
\end{equation}
A Tricomi function $U(a,c,z)$ is an analytical function of its
arguments and for $a=-n$ $(n=0,1,2,\ldots)$ it defines a
polynomial of order $n$ in $z$:
\begin{equation}\label{U-L}
    U(-n,\alpha+1,z) = (-1)^n n!\, L^\alpha_n(z) \ .
\end{equation}
Moreover, using the following property of $U$ (an analog of
(\ref{FeF}) for $_1F_1$)
\begin{equation}\label{}
    U(a,c,z) = z^{1-c}\, U(1+a-c,2-c,z)\ ,
\end{equation}
one obtains
\begin{equation}\label{RR}
    \int_0^\infty \overline{ {\cal R}_{\epsilon\nu}(\varrho)} {\cal
    R}_{\epsilon\nu}(\varrho) \, \varrho d\varrho =
    \frac{\hbar}{2\omega}\,|N_{\epsilon\nu}|^2 \int_0^\infty\, z^{i\nu} \,
    e^{-z} U^2(b,i\nu +1,z)\, dz\ ,
\end{equation}
with $z = \omega\varrho^2/\hbar$.  Now for $b=-n$,  ${\cal
R}_{\epsilon\nu}$ belongs to the Hilbert space $L^2(
\mathbb{R}_+,\varrho d\varrho)$. It implies
\begin{equation}\label{}
\epsilon = \hbar \omega (2n + 1 + i\nu)\ .
\end{equation}
and hence it reproduces discrete spectrum `$\hbar\omega \times
{\rm integer}$' iff $i\nu = l = 0,\pm 1, \pm 2, \ldots\ $.  Now,
using (\ref{U-L}), (\ref{RR}) and
\begin{equation}\label{}
    \int_0^\infty e^{-z}z^\alpha L^\alpha_n(z)L^\alpha_m(z)\, dz =
    \frac {1}{n!}\, \Gamma(n+\alpha+1)\, \delta_{nm}\ ,
\end{equation}
with $\alpha>-1$, one obtains the following  family ${\cal R}_{nl}
\in L^2( \mathbb{R}_+,\varrho d\varrho)$:\footnote{There is a
difference in normalization factor in formulae (37) in
\cite{Bla04}. It follows from slightly different definition of
$L^\alpha_n$.}
\begin{equation}\label{}
{\cal R}_{nl}(\varrho) =
\sqrt{\frac{2\omega/\hbar}{n!\Gamma(n+l+1)}}\
(\sqrt{\omega/\hbar}\varrho)^{l} \exp(
    -\omega\varrho^2/2\hbar)\, L^l_n(\omega\varrho^2/\hbar)\ ,
\end{equation}
with $n=0,1,2,\ldots\ $, and $l=0,1,2,\ldots\ $, satisfying
\begin{equation}\label{}
 \int_0^\infty \overline{ {\cal R}_{nl}(\varrho)} {\cal
    R}_{n'l}(\varrho) \, \varrho d\varrho = \delta_{nn'}\ .
\end{equation}
We stress  that the family ${\cal R}_{nl}$ is defined for $l \geq
0$ only (otherwise it can not be normalized!). Finally, defining
\begin{equation}\label{}
    \phi_{nl}(\varrho,u) = \frac{1}{\sqrt{2\pi}}\,{\cal
    R}_{nl}(\varrho) \, e^{-ul} \ ,
\end{equation}
one has
\begin{equation}\label{}
    \hat{H}\phi_{nl} =  {\cal E}_{nl}\,
    \phi_{nl}\ ,
\end{equation}
with
\begin{equation}\label{enl}
{\cal E}_{nl} = \hbar\omega(2n + l +1) - i\hbar\gamma\, l\ .
\end{equation}
 There is, however, crucial difference between families
$\muu^\pm_{nl}(\rho,\varphi)$ and $\phi_{nl}(\varrho,u)$. The
family $|\muu^\pm_{nl}\>$ corresponds to the poles of
$\psi_{\varepsilon,l}$ from (\ref{psi-el}). No such correspondence
holds for $|\phi_{nl}\>$ and $\psi_{\epsilon\nu}$ from
(\ref{psi-e-nu}). In particular there is no analog of
Theorem~\ref{MAIN} for $|\phi_{nl}\>$. Moreover, ${\cal E}_{nl}$
contrary to $E_{nl}$ from (\ref{Enl}) does not fit the formula for
complex eigenvalues of Feshbach and Tikochinsky \cite{FT} (see
detailed discussion in \cite{I}). It defines simply another family
which is however not directly related to the spectral properties
of the Bateman Hamiltonian.

\section*{Appendix A}
\def\theequation{A.\arabic{equation}}
\setcounter{equation}{0}

\label{Normalization}

To compute $N_{\varepsilon,l}$ in (\ref{R-e}) let us analyze the
quantity $I_\varepsilon= \int_0^\infty
\overline{R_{\varepsilon,l}(\rho)}\,
    R_{\varepsilon,l}(\rho)\, \rho d\rho$. Clearly, this integral
   diverges ($I_\varepsilon=\delta(0)$), however, its structure
enables one the calculation of $N_{\varepsilon,l}$. One has
\begin{equation}\label{A1}
    I_\varepsilon = \frac{1}{2\gamma}\, |N_{\varepsilon,l}|^2 \,
     \int_0^\infty
    z^{|l|}\, _1\! F_1(a,|l|+1,iz)\,  _1\!
    F_1(\overline{a},|l|+1,-iz)\, dz\ ,
\end{equation}
where we defined $z = \gamma \rho^2$. Now, the integral in
(\ref{A1}) belongs to the general class
\begin{equation}\label{}
    J = \int_0^\infty\, e^{-\lambda z}\,
    z^{\mu -1}\, _1\! F_1(\alpha,\mu,kz)\,  _1\!
    F_1(\alpha',\mu,k'z)\, dz\ ,
\end{equation}
given by the following formula  (see Appendix f in \cite{LL}):
\begin{equation}\label{J}
    J = \Gamma(\mu)\, \lambda^{\alpha-\alpha'-\mu}
    (\lambda-k)^{-\alpha}(\lambda-k')^{-\alpha'}\, _2F_1\left(
    \alpha,\alpha',\mu; \frac{kk'}{(\lambda-k)(\lambda-k')}
    \right)\ .
\end{equation}
Using the above formula with $\lambda=0$, $\mu=|l|+1$, $\alpha=a$,
$\alpha'=\overline{a}$ and $k=-k'=i$ one finds
\begin{equation}\label{}
    I_\varepsilon= \frac{1}{2\gamma}\, |N_{\varepsilon,l}|^2 \,
    (-i)^{-a}\, \overline{  (-i)^{-a}}\,
    _2F_1(a,\overline{a},|l|+1;1)\ .
\end{equation}
Finally, noting that
\begin{equation*}\label{}
_2F_1(\alpha,\beta,\gamma;1) = \frac{\Gamma(\gamma)
\Gamma(\gamma-\alpha-\beta)}{\Gamma(\gamma-\alpha)
\Gamma(\gamma-\beta)}\ ,
\end{equation*}
one has
\begin{equation}\label{II}
 I_\varepsilon= \frac{|l|!}{2\gamma}\,  |N_{\varepsilon,l}|^2 \,
    (-i)^{-a}\, \overline{  (-i)^{-a}}\,
    \frac{\Gamma(0)}{\Gamma(a)\Gamma( \overline{a})}\ .
\end{equation}
Therefore, comparing (\ref{II}) with $I_\varepsilon=\delta(0)$ one
finds
\begin{equation}\label{}
N_{\varepsilon,l} = \sqrt{\frac{\gamma}{\pi |l|!}}\,
 (-i)^a\, \Gamma(a)\ ,
\end{equation}
which proves (\ref{N-el}).

\section*{Appendix B}
\def\theequation{B.\arabic{equation}}
\label{Proof} \setcounter{equation}{0}

Due  to the Gel'fand-Maurin spectral theorem \cite{RHS1,RHS2} an
arbitrary function $\phi^+\in \Phi_+$ may be decomposed with
respect to the basis $\psi_{\varepsilon,l}$
\begin{equation} \label{GM-1}
\phi^+ = \sum_{l=-\infty}^\infty \int_{-\infty}^\infty
d\varepsilon\,   \psi_{\varepsilon,l}  \<
\psi_{\varepsilon,l}|\phi^+\> \ .
\end{equation}
Now, since $ \la \psi_{\varepsilon,l} | \phi^+\r \in {\cal
H}^2_+$, we may close the integration contour along the upper
semi-circle $|\varepsilon|\rightarrow \infty$. Applying the
Residue Theorem one obtains
\begin{eqnarray} \label{phi-R}
\phi^+(\rho,\varphi) = 2\pi i \sum_{l=-\infty}^\infty
\sum_{n=0}^\infty \,\mbox{Res}\,
\psi_{\varepsilon,l}(\rho,\varphi)\Big|_{\varepsilon=\varepsilon_{nl}}\,
\la \psi_{\varepsilon,l} |
\phi^+\r\Big|_{\varepsilon=\varepsilon_{nl}} \, .
\end{eqnarray}
Using the well known formula for the residuum
\begin{equation}\label{}
    \mbox{Res}\, \Gamma(a)\Big|_{a=-n} = \frac{(-1)^n}{n!}\ ,
\end{equation}
one obtains
\begin{equation}\label{}
 \mbox{Res}\, \psi_{\varepsilon,l}\Big|_{\varepsilon=\varepsilon_{nl}} =
 \frac{-i}{\sqrt{i^{2n+|l|+1}}}\, \sqrt{\frac{1}{2\pi\hbar}}\,
  \sqrt{\frac{(n+|l|)!}{n!|l|!}}\ {\muu^+_{nl}}\ .
\end{equation}
Moreover,  the analytical function
$\overline{\psi_{\varepsilon,l}}$ computed at
$\varepsilon=\varepsilon_{nl}$ reads:
\begin{equation*}\label{}
  \overline{\psi_{\varepsilon,l}}\Big|_{\varepsilon=\varepsilon_{nl}}
  = i^{n+|l|+1}\,\sqrt{\frac{\gamma}{\pi |l|!}}\,
  (\sqrt{-i\gamma/\hbar}\,\rho)^{|l|}\,
   \exp(i{\gamma}\rho^2/2\hbar)\,
{_1}\!F_1(n+|l|+1,|l|+1,-i\gamma\rho^2/\hbar)\,
\overline{\Phi_l(\varphi)}\ .
\end{equation*}
Due to the well known relation \cite{GR,Morse,AS}
\begin{equation}\label{FeF}
_1F_1(a,b,z) = e^z\, _1F_1(b-a,b,-z)\ ,
\end{equation}
one finds
\begin{equation}\label{}
 \overline{\psi_{\varepsilon,l}}\Big|_{\varepsilon=\varepsilon_{nl}}
  = \sqrt{i^{2n+|l|+1}}\, \sqrt{\frac{\hbar}{2\pi}}\,
  \sqrt{\frac{n!|l|!}{(n+|l|)!}}\ \overline{\muu^-_{nl}}\ .
\end{equation}
and hence the formula (\ref{phi+}) follows. In a similar way one
shows (\ref{phi-}).

\section*{Acknowledgments}

This work was partially supported by the Polish State Committee
for Scientific Research Grant {\em Informatyka i in\.zynieria
kwantowa} No PBZ-Min-008/P03/03.


\begin{thebibliography}{99}


\bibitem{I} D. Chru\'sci\'nski, J. Jurkowski, {\it Quantum damped oscillator I: dissipation and
resonances}, quant-ph/0506007


\bibitem{Bat31} H. Bateman, Phys. Rev. {\bf 38}, (1931) 815.

\bibitem{Dekker81} H. Dekker, Phys. Rep. {\bf 80}, (1981) 1--112.

\bibitem{Vit92} E. Celeghini, M. Rasetti and G. Vitiello, Ann. Phys. (N.Y.), {\bf 215}
(1992) 156--170.


\bibitem{Bla04} M. Blasone and P. Jizba,
Ann.  Phys. (N.Y.), {\bf 312},  (2004) 354--397.


\bibitem{FT} H. Feshbach and Y. Tikochinsky, in: A Festschrift for I.I. Rabi, Trans. New York
Ac. Sc. Ser. 2 {\bf 38}, (1977) 44.

\bibitem{Kemble} E.C. Kemble, Phys. Rev. {\bf 48} (1935) 549

\bibitem{Wheeler} K.W. Ford, D.L. Hill, M. Wakano and J.A.
Wheeler, Ann. Phys. {\bf 7} (1959) 239

\bibitem{Friedman} W.A. Friedman and C.J. Goebel, Ann. Phys. {\bf
104} (1977) 145

\bibitem{Ann1} G. Barton, Ann. Phys. {\bf 166} (1986) 322

\bibitem{Ann2} N.L. Balazs and A. Voros, Ann. Phys. {\bf 199} (1990) 123

\bibitem{Castagnino} M. Castagnino, R.  Diener, L. Lara and G. Puccini,
     Int. Jour. Theor. Phys. {\bf 36} (1997) 2349

\bibitem{Shimbori1} T. Shimbori and T. Kobayashi, Nuovo Cimento B
{\bf 115} (2000) 325





\bibitem{damp2} D. Chru\'sci\'nski, J. Math. Phys. {\bf 45} (2004) 841.

\bibitem{RES-1} S. Albeverio, L.S. Ferreira and L. Streit, eds. {\em Resonances
 -- Models and Phenomena}, Lecture Notes in Physics {\bf 211}, Springer, Berlin, 1984

\bibitem{RES-2} E. Brandas and N. Elander, eds. {\em Resonances},
 Lecture Notes in Physics {\bf 325}, Springer, Berlin, 1989


\bibitem{RHS1} I.M. Gel'fand and  N.Y. Vilenkin, {\em Generalized
Functions}, Vol. IV, Academic Press, New York, 1964.


\bibitem{RHS2} K. Maurin, {\em General Eigenfunction Expansion and
Unitary Representations of Topological Groups}, PWN, Warszawa,
1968.



\bibitem{Bohm-Gadella} A. Bohm and M. Gadella, {\em Dirac Kets,
 Gamov Vectors and Gel'fand Triplets}, Lecture Notes in Physics {\bf 348},
Springer, Berlin, 1989


\bibitem{Bohm} A. Bohm, H.-D. Doebner, P. Kielanowski, {\em
Irreversibility and Causality, Semigroups and Rigged Hilbert
Spaces}, Lecture Notes in Physics {\bf 504}, Springer, Berlin,
1998.





\bibitem{Fluge} S. Fl\"ugge, {\em Practical Quantum Mechanics},
Springer-Verlag, Berlin,  1999.

\bibitem{Kleinert} H. Kleinert, {\it Path Integrals in Quantum Mechanics, Statistics and Polymer
Physics}, 3rd Edition,  World Scientific, London, 2004.













\bibitem{LL} L.D. Landau and E.M. Lifshitz, {\em Quantum Mechanics}, Pergamon, London, 1958


\bibitem{GR} I. Gradshteyn and I. Ryzhik, {\em Table of Integrals,
Series and Products}, Academic Press, 1965

\bibitem{Morse} P.M. Morse and H. Feshbach, {\em Methods of
Theoretical Physics}, McGraw-Hill, New York, 1953

\bibitem{AS} M. Abramowitz and I. Stegun, {\em Handbook of
Mathematical Functions}, Dover Publications, New York, 1972

\bibitem{BE} H. Bateman and A. Erdelyi, {\em Higher Transcendental
Functions}, McGraw-Hill, New York, 1953.

\bibitem{Duren} P.L. Duren, {\em Theory of  ${\cal H}^p$ Spaces},
Academic Press, New York, 1970

\bibitem{Yosida} K. Yosida, {\em Functional Analysis}, Springer,
Berlin, 1978




\end{thebibliography}
\end{document}